# Comments on "Antarctic Automatic Weather Station Program: 30 Years of Polar Observations"

Krzysztof Sienicki
*Chair of Theoretical Physics of Naturally Intelligent Systems, Topolowa 19, 05-807 Podkowa Leśna, Poland*
kris_sienicki@yahoo.com
(24 February, 2013)

Recently Lazzara et al. (2012) presented a review of the technical and scientific progress in deployment, data collection and analysis of the Automated Weather Stations (AWS) in the Antarctic. In the sub-section entitled Science Applications using AWS Observations, the authors' briefly account for several scientific occurrences of meteorological data collected by AWS. In between they make reference to the well known Captain Robert F. Scott *Terra Nova* Expedition 1910–1912 and its meteorological record re-analysis using AWS data by Solomon and Stearns. However, Lazzara's et al. description is incomplete and does not reflect past and current state of knowledge. This comment offers an important correction of Lazzara's et. al. account on that matter.

Firstly, Lazzara's et al. (2012) on page 1532 comment that "Previously, the weather conditions reported by Scott's party were assumed to be typical [*sic*] for the Ross Ice Shelf in late February and early March [1912]." is not factual. Actually, since finding Captain Scott records on 12 Nov. 1912, it was known that the party reported that it was struck by extremely atypical black swan weather events, February 27-March 19, 1912 - Extreme Cold Snap and March 21-29, 1912 Never-Ending Gale. This knowledge was and is based on Captain Scott's (1913) own meteorological record and journal description.

Secondly, more recent studies (Sienicki, 2011) challenge the accuracy of Captain Scott's record and its subsequent analysis by Solomon and Stearns. In particular by using AWS data, artificial neural network simulations and statistical analysis it was shown that the Extreme Cold Snap and Never-Ending Gale reported by Captain Scott were invented meteorological events that never occurred back in 1912.

It was also shown there that Solomon and Stearns analysis was fallacious reasoning involving the Cherry Picking fallacy. On the contrary to the Lazzara et al. (2012) assessment they used "a broader climatological context" Solomon and Stearns 14 years of AWS data at Schwerdtfeger station (-79.904, 169.97) does not represent a climatological approximation (Bryson, 1997).

Extensive analyses (Sienicki, 2013) in between of the results of Solomon and Stearns publication (1999) and Solomon's book (2001) concerning the meteorological record of Captain Scott sheds new light on the *Terra Nova* Expedition as well as on subsequent meteorological data manipulations including data dragging and fabrication.

**ACKNOWLEDGMENTS**
The author appreciates the support of the Antarctic Meteorological Research Center and the Automatic Weather Station Program for surface meteorological data, NSF grant numbers ANT-0636873, ARC-0713483, ANT 0838834, and/or ANT-0944018.